# A novel filter for accurate estimation of fluid pressure and fluid velocity using poroelastography

Md Tauhidul Islam[a], Raffaella Righetti[a,*]

[a]*Department of Electrical and Computer Engineering, Texas A&M University, College Station, Texas, USA-77840*

**Abstract**

Fluid pressure and fluid velocity carry important information for cancer diagnosis, prognosis and treatment. Recent work has demonstrated that estimation of these parameters is theoretically possible using ultrasound poroelastography. However, accurate estimation of these parameters requires high quality axial and lateral strain estimates from noisy ultrasound radio frequency (RF) data. In this paper, we propose a filtering technique combining two efficient filters for removal of noise from strain images, i.e., Kalman and nonlinear complex diffusion filters (NCDF). Our proposed filter is based on a novel noise model, which takes into consideration both additive and amplitude modulation noise in the estimated strains. Using finite element and ultrasound simulations, we demonstrate that the proposed filtering technique can significantly improve image quality of lateral strain elastograms along with fluid pressure and velocity elastograms. Technical feasibility of the proposed method on an in vivo set of data is also demonstrated. Our results show that the CNRe of the lateral strain, fluid pressure and fluid velocity as estimated using the proposed technique is higher by at least 10.9%, 51.3% and 334.4% when compared to the results obtained using a Kalman filter only, by at least 8.9%, 27.6% and 219.5% when compared to the results obtained using a NCDF only and by at least 152.3%, 1278% and 742% when compared to the results obtained using a median filter only for all SNRs considered in this study.

## 1. Introduction

Elevated interstitial fluid pressure (IFP) is a parameter of fundamental importance in cancer initiation, progression and treatment [56, 18] [40, 26]. IFP has been found to induce metastasis in cancers [56, 18] and to be one of the main causes for therapeutic failure in cancers. IFP also affects drug delivery in chemo- and immuno-therapy [40, 26]. Interstitial hypertension created by IFP is one of the major causes of radiation therapy failure [54, 55]. Interstitial fluid velocity (IFV) created by IFP is also an important parameter for drug delivery and a parameter that affects the metastastatic nature of a tumor as it measures fluid transport inside the tumor [32].

To date, there are only a few invasive methods that can be used to estimate IFP in tissues [46, 6]. Dynamic contrast enhanced MRI has been investigated as a possible tool for imaging IFP and IFV in tissues but with limited success [25, 23, 24, 38]. Elastography is an ultrasound imaging modality that aims at non-invasively imaging the strains generated in a tissue in response to a small compression [45, 44]. Poroelastography is a branch of elastography, which images the temporal behavior of fluid-filled materials [51, 50]. Recently, we have proposed a technique that can estimate fluid pressure and fluid velocity from data acquired from an ultrasound poroelastography experiment and mathematically proved their direct link to IFP and IFV [31].

The estimation of fluid pressure and velocity in an ultrasound poroelastography experiment requires differentiation of strains, and differentiation operation amplifies noise [31]. More specifically, the fluid pressure requires time derivative of both the axial and lateral strains and the fluid velocity requires spatial derivative of the fluid pressure. Thus, estimation of fluid pressure and fluid velocity demands highly accurate axial and lateral strain estimates, which are often difficult to obtain from noisy ultrasound RF data. The presence of even small amount of noise on the RF data (i.e., SNR values > 50 dB, which are already unusual in experiments), the strain images could result in highly noisy fluid pressure and fluid velocity estimations due to the requirement of multiple consecutive derivatives [31].

---

*Corresponding author
  Email address: righetti@ece.tamu.edu (Raffaella Righetti)

In order to design optimal filtering methods for strain elastograms, it is important to understand the noise affecting the strain estimates. Strain elastograms typically suffer from two main types of noise, i.e., decorrelation noise and amplitude modulation (AM) noise [14]. The decorrelation noise can be induced by compression, slide-slippage, and other undesirable motion in the sample. AM noise can be caused by a random fluctuation of the RF signal magnitude, which results in the dislocation of the the displacement estimate from the center of the analysis window. Therefore, the gradient of the displacement divided by the spacing of window can produce incorrect strain estimates. As the AM noise strongly correlates with the features in the RF data (and, consequently, B-mode image), the presence of AM noise in the strain elastogram could be easily misinterpreted as micro-structure in anisoechoic regions of the sample [37].

Several methods have been proposed to mitigate the AM noise in the strain elastograms. First, compression of the signal envelope can be used to reduce the amplitude fluctuations, which may shift the estimation locations towards the center of the window [8]. The second method is adaptive stretching [8, 1], which compensates for the strain between windows by stretching the signal to match closely to the true displacement at all points. Although found to be effective in reducing the AM noise in many cases, the adaptive stretching method is computationally expensive [59]. Another prominent method is amplitude modulation correction (AMC), which is based on computing the correct strains by calculating a weight using the amplitude of pre- and post-compression RF data [37]. Although the above-mentioned methods along with others available in the literature work effectively in many practical cases, they are mostly applicable for strains estimated by window-based methods, where two windows (1D or 2D) between pre- and post-compression RF data are matched to obtain the displacement and strain estimates.

Sample-based strain estimation techniques [52, 53, 29] have shown better performance in estimating both the axial and lateral strains in comparison to the window-matching techniques. However, till date, no method has been proposed to reduce AM noise in these methods.

Kalman filtering has been successfully used in a number of different scientific and engineering fields [16, 11], including ultrasound elastography [53, 28, 39, 29]. In [53], the authors proposed to use Kalman filtering to improve strain estimation. While this method has been shown to successfully remove Gaussian noise, it cannot be used to remove multiplicative noise such as AM noise.

Nonlinear diffusion filtering [48] has also been extensively investigated as a powerful technique for denoising medical images because of its superior ability to reduce noise while preserving edge information [13, 12, 19, 7, 22, 21]. Initially introduced in the original paper by Perona and Malik [48], the method is based on representing the time evolution of an image via a nonlinear diffusion equation. The diffusivity of an image pixel is determined by the diffusion coefficient whose value depends on local gradient and a global noise threshold. This results in heavy smoothing in homogeneous regions whereas the edges are sharpened due to large-valued gradients at the edge locations. However, the method is susceptible to stability issues and fails to remove the noise in highly noisy conditions especially at thin edges. Thus, a number of regularization procedures have been proposed in the literature [7, 2, 49]. In other approaches, the diffusion coefficient is modified by considering the local statistics of the image [61, 62, 34]. In [3], NCDF was used to remove multiplicative speckle noise from B-mode images while preserving important features in the image.

Decorrelation along with other noise sources (such as measurement noise, etc.) in the elastographic images have been modeled using additive Gaussian noise models [53, 42]. Thus, elastographic images are, in general, affected by at least two sources of noise additive Gaussian noise and AM-type of noise. In this paper, we derive a strain noise model composed by multiplicative noise model representing AM noise and additive Gaussian noise representing the additional sources of noise affecting strain elastograms. We also propose a combination of Kalman filtering and NCDF to eliminate both additive Gaussian noise and AM noise from axial and lateral strain elastograms. In the first step of the proposed algorithm, the noisy strain estimates are filtered using Kalman filtering with the intent of reducing or eliminating additive Gaussian noise. In the second step, NCDF is used on the output of the Kalman filter to reduce or eliminate multiplicative AM noise from the strain images. NCDF is computationally fast and noise-adaptive. Unlike other diffusion filters, the noise-adaptivity of NCDF ensures correct amount of noise reduction, which prevents the

under- and over-smoothing of the strain images. We statistically evaluate the performance of the proposed filtering method in estimating axial and lateral strains and compared it to the performances of Kalman alone [53], NCDF alone [3] and median filtering alone [10, 42]. We then show that the availability of the proposed filtering technique can dramatically improve estimation of fluid pressure and fluid velocity in poroelastography applications.

## 2. Proposed strain noise model

In ultrasound elastography, local displacements are estimated from pairs of pre- and post-compression RF data, and, from the displacements, strain elastograms are generated using methods such as the least squared error (LSQ) [33]. The resulting strain elastograms are infiltrated with decorrelation, measurement and AM noise. The first two can be modeled as Gaussian additive noise [37]. The AM noise in the strain estimates is created because of the random fluctuations of the amplitude of the RF signals, which results in speckle noise in B-mode images. The amount of AM noise is directly proportional to the amount of random fluctuations of RF signal amplitude. As random fluctuations of RF signal amplitude increase, the AM noise affecting the strain images increases and vice-versa. Therefore, one possible way to model the AM noise is multiplicative noise, similarly to the way speckle noise is modeled in B-mode images. The multiplicative model of AM noise can also be derived mathematically as shown below.

In [37], the authors proposed the following model for the displacement estimates ($\hat{d}_n$) considering only AM noise at $n^{th}$ window in a window based displacement/strain estimation method in 1D

$$\hat{d}_n \approx \frac{\sum_{t=n\Delta t}^{n\Delta t+T} W(t)d(t)}{\sum_{t=n\Delta t}^{n\Delta t+T} W(t)}, \qquad (1)$$

where $d(t)$ is true displacement of the RF signal at a particular location $t$ inside the window, $W(t)$ is the associated weight estimates, $\Delta t$ is the window spacing and $T$ is the window length. $W(t)$ depends on the value of pre- and post-compression RF envelope (B-mode signal) [37].

From eq. (1), we see that the estimated displacement with AM noise can be written as weighted sum of actual displacements inside a window. If this equation is translated to the strain estimates, it can be written as

$$\hat{z}_n \approx \frac{\sum_{t=n\Delta t}^{n\Delta t+T} W(t)tz(t)}{\tau_n \sum_{t=n\Delta t}^{n\Delta t+T} W(t)}. \qquad (2)$$

Therefore, the strain estimates infiltrated with AM noise also become the weighted sum of the actual strains inside the window. It should be noted that eq. (2) is based on the assumption that the displacement and strain can be related using a linear equation, $d(t) = \alpha + zt$ and $\alpha$ is constant and same in $d(t)$ and $\hat{d}_n$ [37]. $\tau_n$ refers to the center of the $n^{th}$ window. This equation can be approximated for a sample-based estimated strain at $i^{th}$ sample (corresponds to center sample of the window) as

$$\hat{z}_i \approx v_i z_i, \qquad (3)$$

where we assumed that $v_i$ is such a weight that

$$v_i z_i = \frac{\sum_{t=n\Delta t}^{n\Delta t+T} W(t)tz(t)}{\tau_n \sum_{t=n\Delta t}^{n\Delta t+T} W(t)}. \qquad (4)$$

Here, $z_i$ is the actual strain at the $i^{th}$ sample. Therefore, the noisy strain estimate $\hat{z}_i$ can be expressed as a multiplication of the actual strain $z_i$ and an unknown factor $v_i$, which can be treated as generated by AM noise.

Based on the above multiplicative noise model for AM noise and the additive noise model for decorrelation and measurement noise, we can write the noisy strain measurement in 2D as

$$\rho_{i,j} = I_{i,j} + g_{i,j}, \qquad (5)$$

where $I_{i,j} = v_{i,j} z_{i,j}$, $z_{i,j}$ is the true strain field, $v_{i,j}$ is the AM noise and $g_{i,j}$ is Gaussian additive noise. $g_{i,j}$ denotes the sum of decorrelation and measurement noise at pixel $(i, j)$.

## 3. Proposed filtering method

### 3.1. Step I - Kalman Filtering

A Kalman filter is applied in the lateral direction of the noisy strain elastogram to obtain a denoised strain image. Assuming $r_{i,j}$ denotes the Gaussian process noise, we have [17]

$$I_{i,j} = I_{i,j-1} + r_{i,j}. \tag{6}$$

Assuming $\bar{I}_{i,j}$ and $\hat{I}_{i,j}$ the a priori estimate of the strain before step $j$ and a posteriori estimate of the strain after step $j$, the update equations can be written for $\bar{I}_{i,j}$ as [53]

$$\hat{I}_{i,j} = \bar{I}_{i,j} + \frac{\hat{q}_{i,j}}{\hat{q}_{i,j} + \sigma_g^2}(\rho_{i,j} - \bar{I}_{i,j}), \tag{7}$$

$$\hat{q}_{i,j} = \left(1 - \frac{\bar{q}_{i,j}}{\bar{q}_{i,j} + \sigma_g^2}\right)\bar{q}_{i,j}, \tag{8}$$

where $\bar{I}_{i,j} = \hat{I}_{i,j-1}$ and $\sigma_g^2$ is the variance of the additive noise $g$. $\hat{q}_{i,j}$ and $\bar{q}_{i,j}$ are the variances of $\hat{I}_{i,j}$ and $\bar{I}_{i,j}$ and can be related through

$$\bar{q}_{i,j} = \hat{q}_{i,j-1} + \sigma_r^2, \tag{9}$$

where $\sigma_r^2$ is the variance of the process noise $r$. $\hat{q}_{i,j-1}$ is assumed as zero for $j = 1$. $\sigma_r^2$ is determined using $\sigma_r^2 = (\mu_{j-1} - \mu_j)^2$, where $\mu_{j-1}$ and $\mu_j$ are the mean values (computed using a Gaussian kernel of standard deviation of 0.6) of strains in square blocks of $3 \times 3$ pixels around samples $(i, j - 1)$ and $(i, j)$, respectively. Here, $\sigma_g^2$ is the variance of the noisy strain values $\rho_{i,j}$ over the entire strain elastogram and it is the same for every pixel in the strain elastogram.

This method can be applied to both the axial and lateral strain elastograms to remove additive Gaussian noise.

### 3.2. Step II - Nonlinear complex diffusion filtering

The general equation for a non linear complex diffusion filter can be written as [58]

$$\frac{\delta I}{\delta t} = \nabla \cdot (D\nabla I), \tag{10}$$

where $I$ is the estimated strain image after Kalman filtering (step 1), $\nabla \cdot$ is the divergence, $\nabla$ is the gradient and $D$ is the diffusion coefficient.

This equation can be discretized by using a finite difference scheme as follow

$$I_{i,j}^{m+1} = I_{i,j}^m + \Delta t^m (\bar{D}_{i,j}^m) \Delta_d I_{i,j}^m + \nabla_d D_{i,j}^m \cdot \nabla_d I_{i,j}^m, \tag{11}$$

where $\Delta_d$ and $\nabla_d$ are the discrete Laplacian and gradient operator, $\Delta t^m$ is time step at the iteration $m$. Here, we assume that after a certain number of iterations, $I_{i,j}^{m+1} \approx z_{i,j}$, which is the true strain field.

The average diffusion coefficient $\bar{D}_{i,j}^m$ at pixel number $(i, j)$ and at iteration $m$ can be written as

$$\bar{D}_{i,j}^m = \frac{4D_{i,j}^m + D_{i+1,j}^m + D_{i-1,j}^m + D_{i,j+1}^m + D_{i,j-1}^m}{8}. \tag{12}$$

Following the work of Salinas et. al [58], the diffusion coefficient $D$ can be written as

$$D = \frac{\exp(i\theta)}{1 + (\frac{\text{Im}(I)}{k\theta})^2}, \tag{13}$$

where $i = \sqrt{-1}$, $k$ is a threshold parameter and $\theta$ is a small angle. In this equation, $D$ does not require any computation of derivative and thus is not ill-posed. As the term $\frac{\text{Im}(I)}{\theta}$ is proportional to the Laplacian of $I$, diffusion is maximized in smooth areas and minimized at the edges of the image.

In eq. (13), $k$ is an important parameter that controls the behavior of $D$ in reducing the noise at edges and smooth regions. We choose $k$ as [3]

$$k = k_{\max} + (k_{\min} - k_{\max}) \frac{g - \min(g)}{\max(g) - \min(g)}, \tag{14}$$

where $k_{max}$ and $k_{min}$ are the maximum and minimum values of $k$ and $min(g)$ and $max(g)$ denote the minimum and maximum value of a function $g$ defined as

$$g = G_{N,\sigma} * Re(I), \tag{15}$$

where $*$ denotes the convolution operation and $G_{N,\sigma}$ is a local Gaussian kernel of $N \times N$ and standard deviation of $\sigma$.

The time step $\Delta t^m$ at iteration $m$ is given by [3]

$$\Delta t^m = \frac{1}{b}(a + (1-a)\exp\left\{\left(-\max\left(|Re(\frac{\delta I^m}{\delta t})|/Re(I^m)\right)\right)\right\}. \tag{16}$$

In this equation, the time step is small at the first iterations because of large time variations in the strain image. At successive iterations, the time step becomes larger as the noise reduces, and changes of the strains become small (fraction-wise) over time. Values of $a$ and $b$ control the dynamic range of the time step.

## 4. Simulations

*4.0.1. Finite element simulations*

The commercial finite element modeling software Abaqus, Dassault Systemes Simulia Corp., Providence, RI, USA was used to validate the technique proposed in this paper. An 'effective stress' principle is used in ABAQUS [27], whereby the total stress acting at a point is assumed to be made up of the average pressure in the pore fluid and an 'effective/elastic stress' on the solid matrix. Both the inclusion and background of the simulated samples were modeled as composed by a linearly elastic, isotropic, permeable solid phase fully saturated with fluid.

The interstitial permeability of the sample was assumed to be independent of the strain and void ratio. The mesh used to model the sample was CAX4RP with 63, 801 elements in the solution plane. The dimension of the solution plane of the sample was 2 cm in radius and 4 cm in height. The inclusion is of 0.75 cm in radius. A zero fluid pressure boundary condition on the right hand side of the sample was imposed. The specific weight of the fluid was assumed to be 1 Nm$^{-3}$ to match the definitions of interstitial permeability in ABAQUS and in the developed model. Under the assumption of unit specific weight of the pore fluid, the hydraulic conductivity and permeability become equal [4]. The void ratio used in all samples was 0.4. To simulate a poroelastography experiment, an instantaneous load of 3700 Pa was applied to the sample in 0.01 second and then kept constant for 360.01 s while the sample was under compression. The time response of each sample was recorded for 360 s with a 1 s sampling interval.

From a mechanical point of view, in our model, the background is simulated to resemble normal tissue while the inclusion is simulated to resemble a tumor. Thus, the mechanical properties of the sample used in our simulations were chosen based on the data available in the literature [36, 43]. Table 1 provides a summary of the values used for the simulation study. The Poisson's ratio of soft tissues (both tumors and normal tissues) is reported in the literature to have a range of values between $0.2 - 0.49$ [36, 60, 41, 20]. In our study, we assumed the Poisson's ratio to be 0.4 in the background tissue and 0.3 in the inclusion. The Young's modulus of the background was assumed to be 32.78 kPa [36]. Tumors have been reported to have a range of Young's modulus contrast (1.1-20) with respect to normal tissues [63, 57]. Therefore, we choose the Young's modulus of the tumor as 70.47 kPa, which is 2.15 times higher than that of normal tissue. The interstitial permeability of the normal tissue was always assumed to be 100 times higher than the interstitial permeability of the tumor. Similar values of interstitial permeability contrast between tumor and surrounding tissue have been previously considered in the literature [60] (supplementary p. 17).

*4.0.2. Ultrasound simulations*

The simulated pre- and post-compression temporal ultrasound radio frequency (RF) data were generated from the mechanical displacements obtained from FEM using a convolution model [15]. Bilinear interpolation was performed on the input mechanical displacement data prior to the computation of the simulated RF frames [9]. The simulated ultrasound transducer resembled our experimental system with 128 elements, frequency bandwidth between $5 - 14$ MHz, a 6.6 MHz center frequency, and 50% fractional bandwidth at $-6$ dB. The transducer's beamwidth was assumed to be dependent on the wavelength and to be approximately 1 mm at 6.6 MHz [51]. The sampling frequency was set at 40 MHz and Gaussian noise was added to set the SNR at different levels (30-60 dB). To compute the axial and lateral strain elastograms from ultrasound pre- and post-compressed RF data, the method proposed in [29] was used.

Table 1: Mechanical parameters of sample A

| Sample name | $E_b$ (kPa) | $E_i$ (kPa) | $\nu_b$ | $\nu_i$ | $k_b$ (m$^4$N$^{-1}$s$^{-1}$) | $k_i$ (m$^4$N$^{-1}$s$^{-1}$) | $\chi_b$ (Pa s)$^{-1}$ | $\chi_i$ (Pa s)$^{-1}$ |
|---|---|---|---|---|---|---|---|---|
| A | 32.78 | 70.47 | 0.4 | 0.3 | $3.19 \times 10^{-10}$ | $3.19 \times 10^{-12}$ | $1.89 \times 10^{-9}$ | $5.67 \times 10^{-9}$ |

Fluid pressure and fluid velocity were estimated from the axial and lateral strain elastograms using the theoretical models reported in [31]. In summary, the fluid pressure at time $t$ can be written as [31]

$$p(R,t) = -K(\epsilon(R,t) - \epsilon(R,\infty)), \qquad (17)$$

where $K$ is the compression modulus of the tumor and $\epsilon(R,t)$ is the volumetric strain at radial position $R$ and time $t$. Estimation of the fluid pressure inside the tumor requires estimation of the Young's modulus and Poisson's ratio, which was performed according to the method described in [30]. The permeability-normalized fluid velocity with respect to the solid along the radial direction inside a tumor can be written as [31]

$$v_R(R,t) = -\frac{dp(R,t)}{dR}. \qquad (18)$$

## 5. Experiments

We report in vivo data obtained from one mouse animal model with triple negative breast cancer [47]. In vivo data acquisition was approved by the Houston Methodist Research Institute, Institutional Animal Care and Use Committee (ACUC-approved protocol # AUP-0614-0033). Details about the in vivo experiment have been reported previously [29].

## 6. Configuration of the filtering methods

In the proposed technique, the length of the Kalman window ($W_k$) was taken as 13 for estimating both the axial and lateral strains. The value of $\theta$ was taken as $\frac{\pi}{30}$ in eq. (13). $k_{max}$ and $k_{min}$ were taken as 28 and 2 in eq. (14). In eq. (15), the value of $N$ was taken as 3 and of $\sigma$ as 10. In eq. (16), the value of $b$ was taken as 4 and $a$ was taken as 0.25. Values of all the parameters used in the proposed method are tabulated in Table 2. The same values of parameters used in step-II of the proposed method were used in the case of NCDF only method. In the Kalman filter only technique, the length of Kalman window was taken as 43 for axial strain estimation [53] and 13 for the lateral strain. In the median filter technique, a median filter of size $5 \times 5$ pixels was applied to the displacement image of $128 \times 128$ pixels and then the least square technique was used for strain estimation from the filtered displacements [10, 42].

## 7. Image quality analysis

Image quality of strain, fluid pressure and fluid velocity elastograms was quantified using two elastographic quality factors, which are typically used for elastographic studies: CNRe (elastographic contrast-to-noise ratio) and PRE (percent relative error). Definition of the CNRe is given in [5].

The PRE is defined as [35]

$$\text{PRE} = \sum_{r=1}^{R} \sum_{c=1}^{C} \frac{\tau_e(r,c) - \tau_t(r,c)}{\tau_t(r,c)} \times \frac{100}{R \times C}, \qquad (19)$$

where $\tau_e$ is the estimated poroelastography parameter (i.e., any one of lateral strain, fluid pressure, fluid velocity) and $\tau_t$ is the true, ideal parameter from FEM; $R$ and $C$ are the number of row and column of $\tau_e$.

Table 2: Values of parameters used in the proposed method

| Parameter | Value |
|---|---|
| $W_k$ | 13 |
| $\theta$ | $\frac{\pi}{30}$ |
| $k_{max}$ | 28 |
| $k_{min}$ | 2 |
| N | 3 |
| $\sigma$ | 10 |
| a | 0.25 |
| b | 4 |

## 8. Results

*8.1. Simulations*

The lateral strain at time point of 36 s estimated using the various filtering methods are shown in Fig. 1 along with the FEM results. We note that the lateral strain elastogram (B1) estimated using Kalman only is less noisy than the results obtained using the NCDF and median filter only. However, granular-type noise (associated to AM noise) can be clearly seen both inside and outside the inclusion in these images. The lateral strain computed by the proposed method (shown in Fig. 1 (B3)) has the highest resemblance with the FEM image. It appears smoother but, at the same time, blurrier than those obtained from the other three techniques.

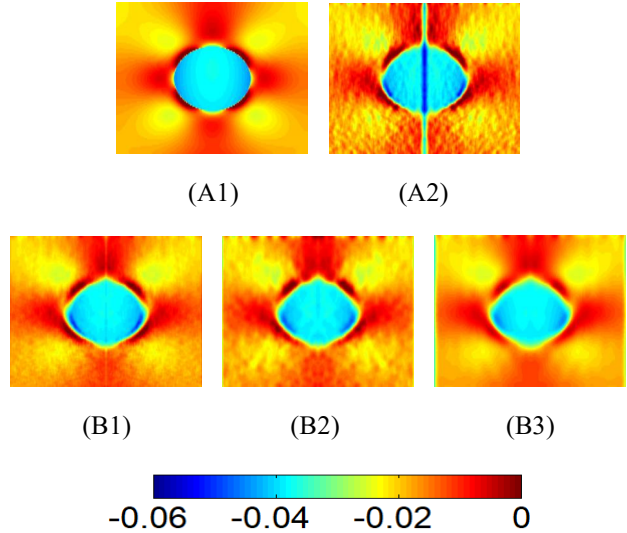

Figure 1: (A1) True lateral strain at time point of 36 s, estimated lateral strains at the same time point at input SNR of 60 dB by (A2) median filter (B1) Kalman filter (B2) NCDF and (B3) proposed technique.

The efficacy of the proposed method in reducing the noise affecting the lateral strain elastogram can be seen more evidently in the zoomed lateral strain image in the background region as shown in Fig. 2 (A2, A3, B1, B2, B3). We see, that outside the inclusion, the granular-type noise is significantly reduced in the proposed method's result only.

Similar observations are also seen for the zoomed lateral strains estimated by different techniques at the inclusion part, which are shown in Fig. 3 (A2, A3, B1, B2, B3). By visual inspection of these results, we see that the proposed technique can more accurately estimate the strain patterns inside the inclusion with respect to the other three methods.

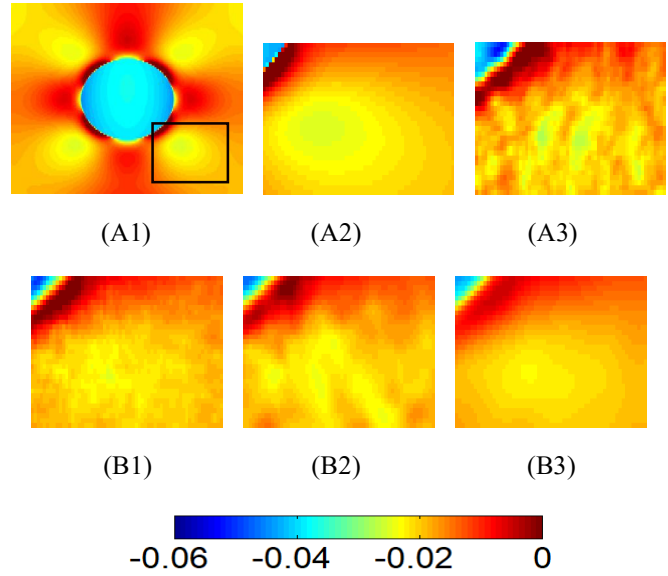

Figure 2: (A1) True lateral strain elastogram at 36 s, zoomed part from the right bottom portion of (A2) true lateral strain elastogram, estimated lateral strain elastogram at input SNR of 60 dB by (A3) median filter (B1) Kalman filter (B2) NCDF and (B3) proposed technique.

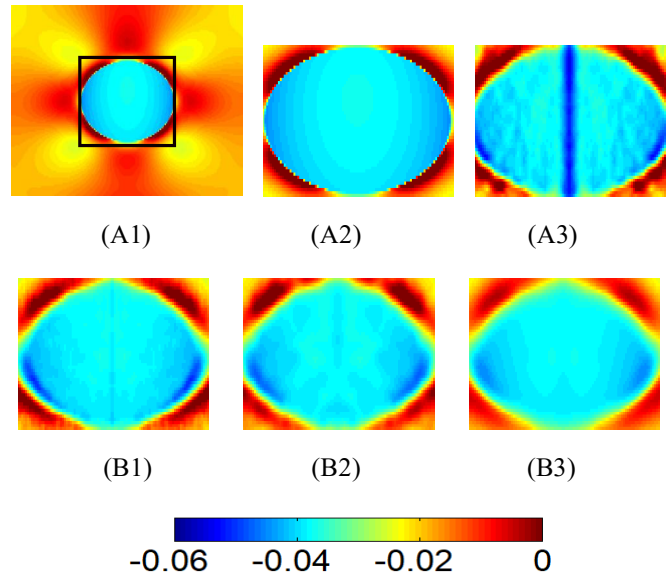

Figure 3: (A1) True lateral strain elastogram at 36 s, zoomed part from the center portion of (A2) true lateral strain elastogram, estimated lateral strain elastogram at input SNR of 60 dB by (A3) median filter (B1) Kalman filter (B2) NCDF and (B3) proposed technique.

For the axial strain images, the improvement obtained using the proposed technique is not as evident as for the lateral strain images but still visually appreciable as shown in Figs. 4 and 5.

The fluid pressure elastograms at three different time points (36 s, 108 s and 180 s) estimated using axial and lateral strain elastograms after being filtered using the different techniques are shown in Fig. 6 along with the FEM results. We see from this figure that the results using NCDF have some similarities with the FEM results but are overall noisy. Kalman filtering results appear better than those from median filtering but are overall very noisy. The fluid pressures using the proposed technique are the only one clearly depicting the pattern of true fluid pressure as shown in the FEM results. Specifically, the theoretically expected 2D bell shaped profile of the fluid pressure from the center to the periphery of the lesion can only be appreciated in Fig. 6 (E2), which matches well with the FEM fluid pressure at the same time point.

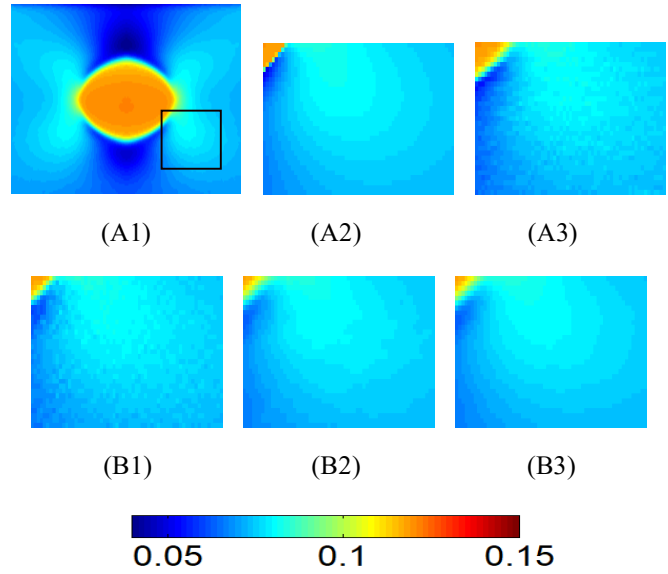

Figure 4: (A1) True axial strain elastogram at 36 s, zoomed part from the right bottom portion of (A2) true axial strain elastogram, estimated axial strain elastogram at input SNR of 60 dB by (A3) median filter (B1) Kalman filter (B2) NCDF and (B3) proposed technique.

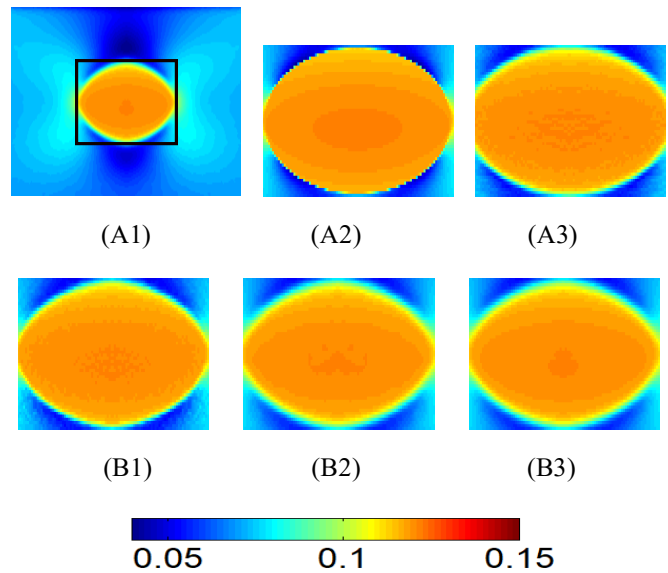

Figure 5: (A1) True axial strain elastogram at 36 s, zoomed part from the center portion of (A2) true axial strain elastogram, estimated axial strain elastogram at input SNR of 60 dB by (A3) median filter (B1) Kalman filter (B2) NCDF and (B3) proposed technique.

The importance of using our proposed technique in reducing noise in axial and lateral strain elastograms prior computation of fluid pressure and fluid velocity is also demonstrated by the results shown in Fig. 7. From this figure, we see that the fluid velocities obtained using the proposed technique are the only ones visually resembling the FEM results.

The computed statistical factors, i.e., CNRe and PRE, for the estimated lateral strain, fluid pressure and fluid velocity at a given time point (36 *s*) obtained by the different techniques are shown in Figs. 8 and 9 for different input

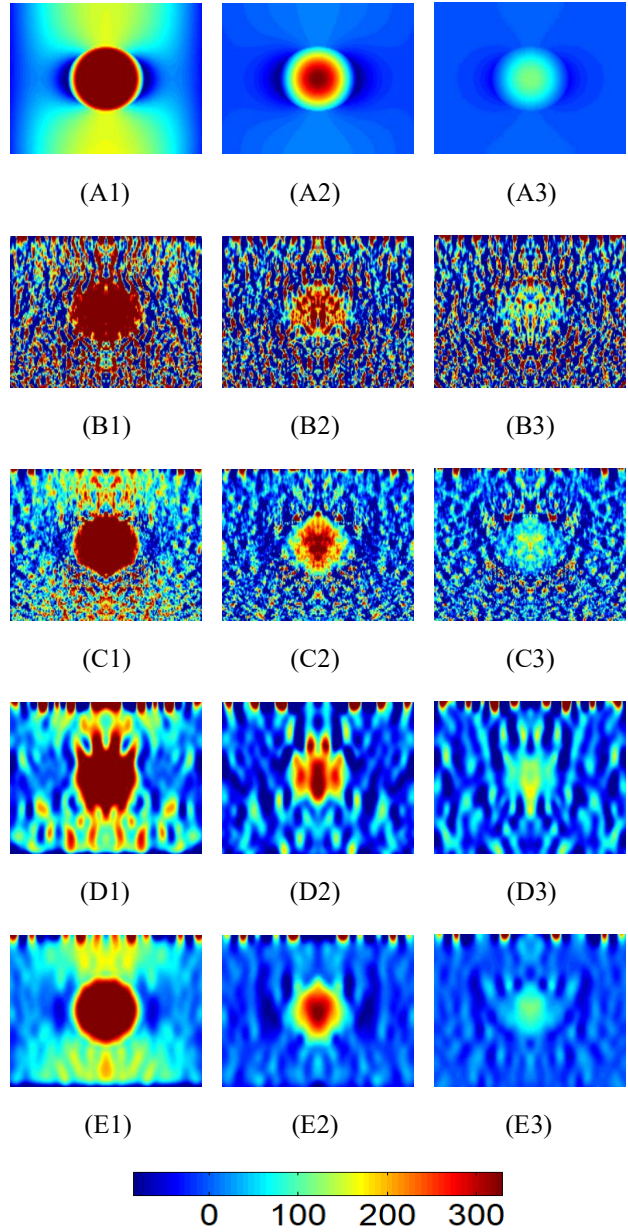

Figure 6: (A) Fluid pressure (in Pa) at different time points (36 s (1), 108 s (2) and 180 s (3)) from FEM, estimated fluid pressure at input SNR of 60 dB by (B) median filter, (C) Kalman filter (D) NCDF (E) proposed method.

SNR levels. We see that the CNRe for all methods increases for increased input SNR. However, the CNRe associated to the proposed technique is the highest across all SNRe levels. The advantages of using the proposed method are particularly evident when observing the fluid pressure and fluid velocity estimation data, where the NCDF, Kalman and median filtering show level of CNRe in the range 0-1 especially for relatively low input SNR values.

The computed PREs for lateral strain, fluid pressure and fluid velocity inside the inclusion estimated by the different techniques are shown in Fig. 9 (A, B, C). In terms of lateral strain estimation, the proposed technique, NCDF and Kalman filtering show similar results. However, in terms of fluid pressure and fluid velocity, Fig. 9 (B) and Fig. 9 (C), the proposed technique performs significantly better than the other three techniques in lower SNR levels. This is especially evident for the fluid velocity data, where we see that the fluid velocity by the proposed technique has PRE of around $10 - 15\%$, whereas PRE in the estimated fluid velocity by other techniques can be $100 - 300\%$.

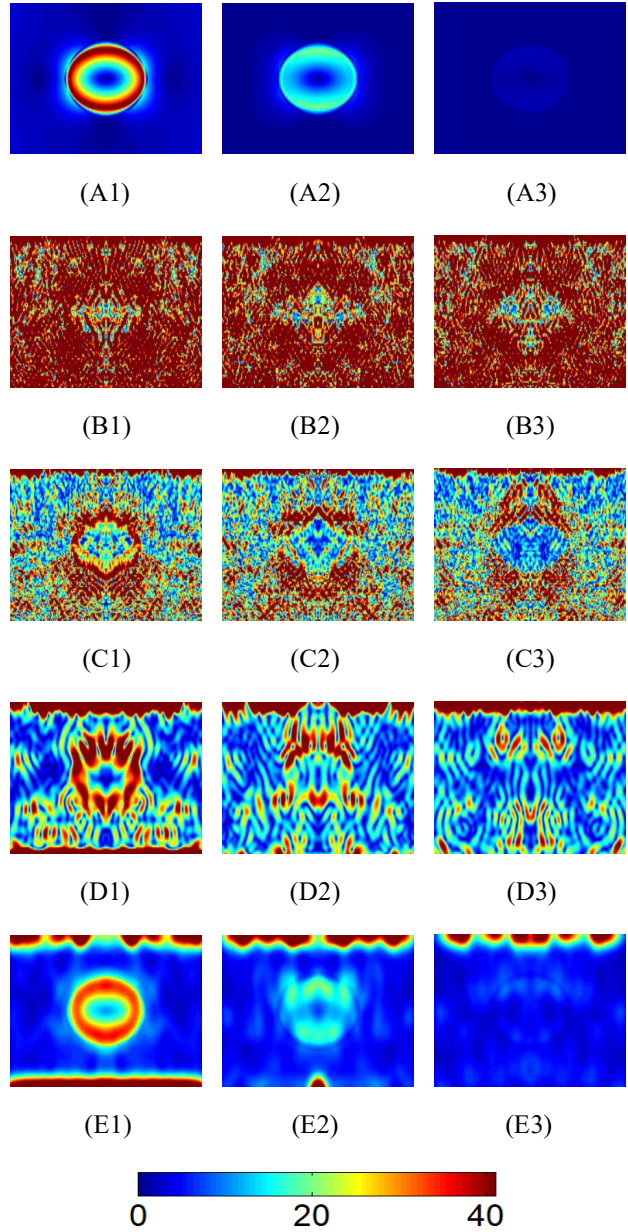

Figure 7: (A) Fluid velocity (in Pa pixel$^{-1}$) at different time points (36 s (1), 108 s (2) and 180 s (3)) from FEM, estimated fluid velocity at input SNR of 60 dB by (B) median filter, (C) Kalman filter, (D) NCDF and (E) proposed method.

*8.2. Experiments*

In vivo lateral strain, axial strain, fluid pressure and fluid velocity elastograms obtained using the various filtering techniques obtained from a cancer animal model are shown in Fig. 10. We see from this figure that the strains from median filter technique are noisier than those obtained using the other methods. Visually, the axial and lateral strains from Kalman filter, NCDF and the proposed method look similar. However, on deeper inspection, we can observe that the strain elastograms resulting from the proposed technique are smoother than those obtained from NCDF and Kalman filter and do not appear affected by granular noise. This noise gets severely amplified in the computation of the fluid pressure and fluid velocity elastograms as shown by these results. The fluid pressure elastograms from NCDF, Kalman and median filters are much noisier than that obtained from the proposed method. The bell-shaped distribution of the fluid pressure can in fact be clearly seen only in the fluid pressure image estimated by the proposed method. The fluid velocity results are similar to the fluid pressure results as only the proposed method allows a clear visualization of the fluid velocity in the tumor.

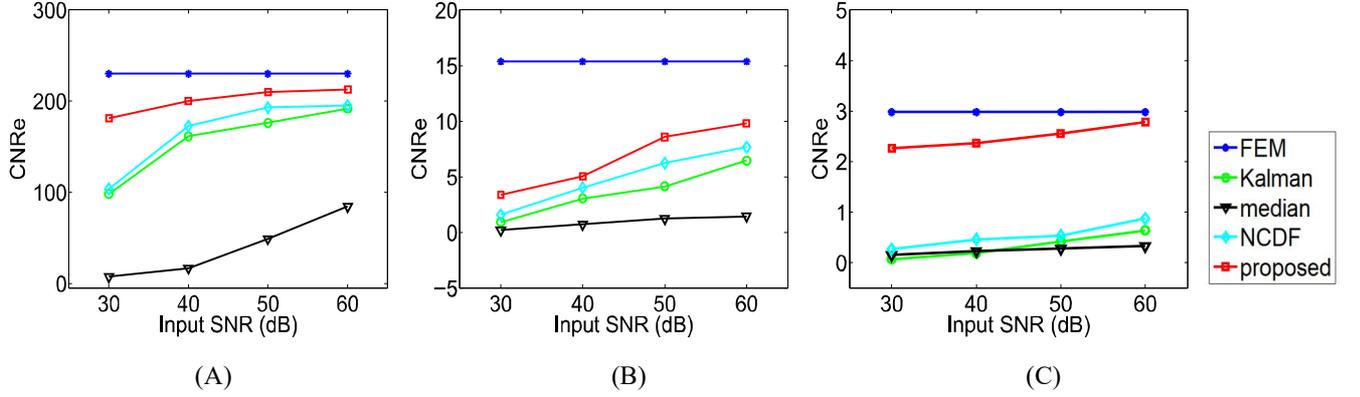

Figure 8: CNRe in estimated lateral strain (A) and fluid pressure (B) and fluid velocity (C) by different filtering techniques at different input SNRs.

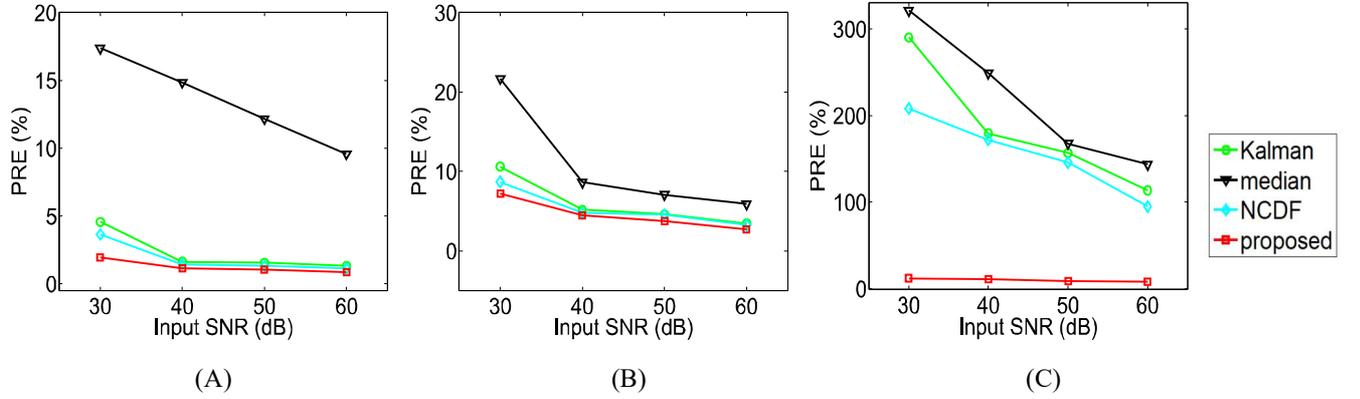

Figure 9: PRE in estimated lateral strain (A) and fluid pressure (B) and fluid velocity (C) by different filtering techniques at different input SNRs.

## 9. Discussion

Fluid pressure and fluid velocity carry important information about the microenvironment of tumors. In previous works, we have theoretically demonstrated that fluid pressure and fluid velocity can be estimated using pororelastography by knowledge of the temporal axial and lateral strain elastograms. Since computation of fluid pressure and fluid velocity requires differentiation of the strain images, to accurately estimate these parameters, it is important to use high quality strain images. In this paper, we have proposed a method, which is based on the combination of Kalman filtering and NCDF. The idea behind this two-step filtering technique is to be able to eliminate both additive and AM noise from the strain images. Our results show that the newly proposed technique is able to produce high quality strain images as well as fluid pressure and fluid velocity images.

As for any filtering technique, the parameter selection for the proposed technique is very important. In our study, the value of the Kalman window was chosen empirically in such a way that it could adequately remove the additive noise with sacrifice of minimum spatial resolution. A value that is too large can result in strain images with lower spatial resolution. The time step in the diffusion filter is also very important. This step needs to be sufficiently small to allow the progress of the diffusion smoothly so that the iteration can be stopped at a time when the noise in the strain image is very low but further diffusion can cause over-smoothing. However, extremely small values can result in very high computational times to reach a solution.

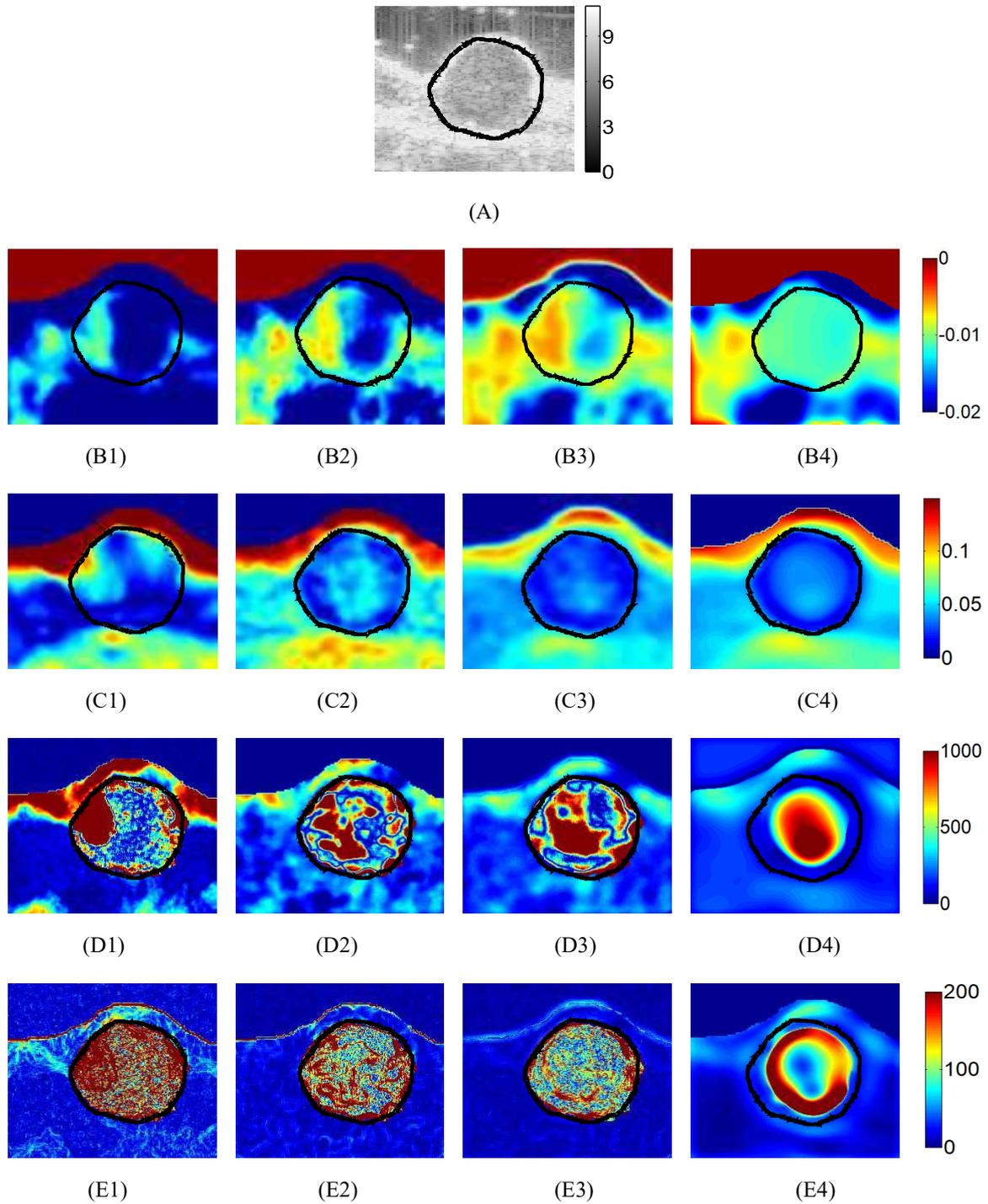

Figure 10: (A) B-mode image, (B) lateral strain (C) axial strain (D) fluid pressure (in Pa) and (E) fluid velocity (in Pa pixel$^{-1}$) estimated by (1) median filter (2) Kalman filter (3) NCDF and (4) proposed method from in vivo experiment at time point of 2 s.

In this paper, the proposed method has been used in combination with the strain estimation technique proposed in [29]. However, it is expected to be useful also when used in conjunction with other sample-based strain estimation techniques [52, 53].

As for most filtering methods used to reduce noise in medical images, the application of the proposed filter is accompanied with a reduction of the spatial resolution of the strain elastograms. In general, the idea is to select filtering parameters such as a compromise between contrast-to-noise ratio and spatial resolution is reached, which is typically application-dependent. Another limitation of the proposed technique is the computational speed since the proposed method requires two different filtering techniques. The proposed technique requires 0.96 s to generate one lateral strain image in a 3.8 GHz Core i5 CPU with 8 GB RAM, whereas median, Kalman and NCDF techniques require 0.05 s, 0.1 s, 0.79 s, respectively. However, running the algorithms in multi-processors can improve the computation speed of the method in the future.

## 10. Conclusion

In this paper, we propose a combination of Kalman and nonlinear complex diffusion filters to remove both additive and AM noises from noisy strain images obtained in ultrasound elastography. Through finite element and ultrasound simulations, we prove the efficacy of the proposed technique in accurate estimation of the axial and lateral strains, fluid pressure and fluid velocity from an ultrasound poroelastography experiment. Clinical feasibility of the technique is also demonstrated in an in vivo animal model tumor.

## 11. Acknowledgment

None.